\documentstyle[preprint,aps]{revtex}

\tighten
\begin{document}
\draft
\preprint{TPI-MINN-00/24 $\;\;$
          UMN-TH-1902-00 $\;\;$
          ITP-SB-00-20   $\;\;$
          }

\newcommand{\nc}{\newcommand}
\nc{\vivi}{very interesting and very important}
\nc{\al}{\alpha}
\nc{\ga}{\gamma}
\nc{\de}{\delta}
\nc{\ep}{\epsilon}
\nc{\ze}{\zeta}
\nc{\et}{\eta}
\newcommand{\th}{\theta}
\nc{\tmn}{\theta_{\mu\nu}}
\nc{\Th}{\Theta}
\nc{\ka}{\kappa}
\nc{\la}{\lambda}
\nc{\rh}{\rho}
\nc{\si}{\sigma}
\nc{\ta}{\tau}
\nc{\up}{\upsilon}
\nc{\ph}{\phi}
\nc{\ch}{\chi}
\nc{\ps}{\psi}
\nc{\om}{\omega}
\nc{\Ga}{\Gamma}
\nc{\De}{\Delta}
\nc{\La}{\Lambda}
\nc{\Si}{\Sigma}
\nc{\Up}{\Upsilon}
\nc{\Ph}{\Phi}
\nc{\Ps}{\Psi}
\nc{\Om}{\Omega}
\nc{\ptl}{\partial}
\nc{\del}{\nabla}
\nc{\be}{\begin{equation}}
\nc{\ee}{\end{equation}}
\nc{\bea}{\begin{eqnarray}}
\nc{\eea}{\end{eqnarray}}
\nc{\ov}{\overline}
\nc{\gsl}{\!\not}
\newcommand{\s}{\mbox{$\sigma$}}
\newcommand{\bi}[1]{\bibitem{#1}}
\newcommand{\fr}[2]{\frac{#1}{#2}}
\newcommand{\gm}{\mbox{$\gamma_{\mu}$}}
\newcommand{\gn}{\mbox{$\gamma_{\nu}$}}
\newcommand{\Le}{\mbox{$\fr{1+\gamma_5}{2}$}}
\newcommand{\R}{\mbox{$\fr{1-\gamma_5}{2}$}}
\newcommand{\GD}{\mbox{$\tilde{G}$}}
\newcommand{\gf}{\mbox{$\gamma_{5}$}}
\newcommand{\Ima}{\mbox{Im}}
\newcommand{\Rea}{\mbox{Re}}
\newcommand{\Tr}{\mbox{Tr}}
\newcommand{\psl}{\slash{\!\!\!p}}
\newcommand{\cp}{\;\;\slash{\!\!\!\!\!\!\rm CP}}
\newcommand{\qq}{\langle \ov{q}q\rangle}
\def\ga{\mathrel{\raise.3ex\hbox{$>$\kern-.75em\lower1ex\hbox{$\sim$}}}}
\def\la{\mathrel{\raise.3ex\hbox{$<$\kern-.75em\lower1ex\hbox{$\sim$}}}}

\title{Low-energy Limits on the Antisymmetric Tensor Field 
Background on the Brane and on the Non-commutative Scale}

\vspace{1cm}

\author{Irina Mocioiu$^1$\footnote{mocioiu@insti.physics.sunysb.edu},
Maxim Pospelov$^2$\footnote{pospelov@mnhepw.hep.umn.edu} 
          and Radu Roiban$^{1}$\footnote{
roiban@insti.physics.sunysb.edu}}

\vspace{1cm}

\address{$^1$ C.N. Yang Institute for Theoretical Physics\\
State University of New York, Stony Brook, NY 11794-3840\\ $\;$\\
$^2$ Theoretical Physics Institute, School of Physics and Astronomy \\
         University of Minnesota, 116 Church St., Minneapolis, MN
         55455, USA
         }
\date{\today}

\maketitle

\begin{abstract}

A non-vanishing vacuum expectation value for an antisymmetric 
tensor field
leads to the violation of Lorentz invariance on the brane. This
violation is  controlled by the $\theta _{\mu \nu }$ parameter, 
which has a
dimension of inverse mass squared. We assume that the 
zeroth order term in $%
\theta $-expansion represents the Standard Model and study the effects
induced by linear terms in $\theta _{\mu \nu }$. Low-energy precision
experiments place the limit on the possible size of this background 
at the
level of $1/\sqrt{\theta }\mathrel{\raise.3ex\hbox{
$>$\kern-.75em\lower1ex%
\hbox{$\sim$}}}5~\times 10^{14}$ GeV. This poses certain 
difficulty for the
TeV-range string scale models, in which the antisymmetric field has to
acquire a relatively large mass to ensure that $\theta _{\mu \nu }=0$ and
avoid cosmological moduli problem.
\end{abstract}

\vspace{4cm}

\hfill\eject

\section{Introduction}

One of the initial motivations for field theories on 
non-commutative spaces was their intrinsically more convergent behavior in 
the ultraviolet regime than the one observed for ordinary field theories.

Field theories on noncommutative spaces (NCFT) can be defined 
as theories in their own right,
independent of string theory. The coordinates in these 
spaces are represented by
self-adjoint operators acting on some Hilbert space ${\cal H}$ and satisfying
the following commutation relations:
\be
[{\hat x}^\mu,\,{\hat x}^\nu]=i\theta^{\mu\nu}~~~~~~~~
[\theta^{\mu\nu},\,x^\rho]=0
\ee
Consequently, fields on such a space are replaced by operators. 
To each such operator
one can associate an ordinary field on a commutative space as follows:
\be
\phi(x)={1\over (2\pi)^{d/2}} \int d^dk {\rm e}^{i k_{\mu}x^{\mu}} 
\Tr [{\hat \phi}({\hat x}) {\rm e}^{-i k_{\mu} {\hat x}^{\mu}}]
\label{eq:weyl}
\ee
where the trace is taken in the Hilbert space ${\cal H}$. By $\phi$ here we 
denote a generic field; we can associate to it some space-time indices
as it will be the case for gauge fields or fermions.

An action defined on ${\cal H}$ must be naturally 
writable in terms of traces; furthermore, we want
that in the limit $\theta^{\mu\nu}\rightarrow 0$ the 
expression reduces to an ordinary action 
on an ordinary, commutative space. Then, the generic form of the action is
\be
S=\Tr[\,(\theta^{-1}_{\mu\nu} [{\hat x}^\nu,\,{\hat \phi}({\hat x})])^2 
+ P({\hat\phi})]
\ee
where $P$ is some polynomial in ${\hat \phi}({\hat x})$. It is 
not difficult to see that,
using (\ref{eq:weyl}), the commutation relations for ${\hat x}$ and 
the Baker-Campbell-Hausdorff formula,
this action reduces to:
\be
S=\int d^D x\, (\partial_\mu \phi(x))^2 + P_*({\phi})
\label{eq:action}
\ee
where by $P_*({\phi})$ we mean that in $P({\hat\phi})$ we replace 
${\hat \phi}$ by $\phi$
and the product of fields is the Moyal product, given by:
\be
\phi_1*\phi_2 (x)=e^{i\frac{1}{2}\theta^{\mu\nu}\frac{\partial}{\partial\xi^\mu}
\frac{\partial}{\partial\zeta^\nu}} \phi_1(x+\xi)\phi_2(x+\zeta)
|_{\xi=\zeta=0}
\label{eq:starprod}
\ee

Noncommutative field theories became popular among string
theorists with the work of 
Connes, Douglas and Schwarz \cite{cds97} who argued that 
M-theory in constant 3-form background is 
equivalent to the supersymmetric Yang-Mills theory, defined
on a non-commutative torus. 
A second wave of interest was generated by the 
paper of Seiberg and Witten \cite{SW99} which summarized and 
extended earlier ideas about the appearance
of noncommutative geometry in string theory with constant 
NS-NS $B$ field background. 

NCFT are constructed starting from string theory in much 
the same way as the usual field theories.
In particular, one computes the string theory S-matrix 
elements and writes a low energy effective 
action
that reproduces them at the tree level. Furthermore, 
the only difference from the usual 
computation is
that the world sheet propagator is modified by the presence 
of $B$. If the world sheet 
has no boundaries, 
then a constant $B$ field can be gauged away. Thus, a 
constant B flux manifests 
itself only in the presence of world sheet boundaries, 
i.e. in the presence of D-branes. Moreover, 
the same 
argument shows that only the components of $B$ 
parallel to the D-brane can be non-zero. The world 
sheet 
propagator restricted to the boundary is modified by the addition 
of a term 
${i\over 2}\theta^{\mu\nu}\epsilon(\tau-\tau')
$, 
where $\epsilon(\tau-\tau')$ is the step function and 
$\theta^{\mu\nu}=-(2\pi\alpha')^2\left({1\over g+2\pi\alpha' B}
B{1\over g-2\pi\alpha' B}\right)^{\mu\nu}$.
Since all vertex operators contain factors of the type $exp(ik\cdot x)$, 
it is easy to see
that all correlation functions will get the extra factor
\be
{\sl e}^{{i\theta^{\mu\nu}
\sum\limits_{\scriptscriptstyle{i<j=1}}^n k^i_\mu k^j_\nu}} 
\ee
which is just the $*$-product defined in (\ref{eq:starprod}) 
written in momentum space.
Thus, the effective action in the presence of 
B-field has the interpretation of a field theory 
on a noncommutative space with noncommutativity 
parameter given by $\theta$.

Transition from the open string theory to the non-commutative field theory 
becomes explicit in a zero slope limit, $\alpha'\sim 
\epsilon^{1/2}\rightarrow 0$, $g_{\mu\nu}\sim \epsilon
\rightarrow 0$. For example, in this limit, the dynamics of Yang-Mills 
fields living on the brane is
governed by the following action:
\be
S={1\over 2g_{YM}^2}\int Tr_{U(N)}{F}^{nc}_{\mu\nu}*{F}^{nc,\mu\nu}
\label{eq:YMaction}
\ee
where we have taken the open string metric to be $\eta_{\mu\nu}$ and 
${F}^{nc}_{\mu\nu}$ is given by:
\be
{F}^{nc}_{\mu\nu}=\partial_\mu A^{nc}_\nu-\partial_\nu A^{nc}_\mu + 
i A^{nc}_\mu*A^{nc}_\nu- i A^{nc}_\nu*A^{nc}_\mu.
\ee 
Comparing this with equation (\ref{eq:action}) 
we see that this can be interpreted as 
the Yang-Mills action on a noncommutative space 
with noncommutativity parameter $\theta$.
Up to now the structure of $B_{\mu\nu}$ in the directions 
parallel to the brane was not really important. It should be noted, however,
that large electric-like background $B_{0i}$ creates various problems in the 
zero slope limit \cite{problem}, \cite{problemsol}. A solution to these 
problems was proposed in \cite{problemsol} and it leads to (\ref{eq:YMaction}).

%







It became a colloquial wisdom that the parameter $\theta _{\mu \nu }$ does
not necessarily have to be of the order of the inverse Plank scale squared 
\cite{SW99,Madore}. For example, $\theta _{\mu \nu }$ could be 
significantly larger in
the ``brane-world'' proposal \cite{ADD}, in which the string scale $M_{s}$
is much smaller than four-dimensional Plank scale due to the large volume of
extra dimensions, $M_{{\rm Pl}}^{2}=M_{s}^{n+2}V_{n}$. In this case a
``natural'' scale for $\theta _{\mu \nu }$ could be $M_{s}^{-2}$. 
Another
example is the open string realization of the non-commutative field theories
on the brane in the zero slope limit described above, 
which presumes the non-commutative
scale to be fixed, while gravity is decoupled (i.e. $M_{{\rm Pl}}\rightarrow
\infty $). If instead we choose to fix the gravitational scale, zero slope 
limit would correspond to a {\em large} non-commutative parameter in units of 
$M_{{\rm Pl}}^{-2}$. 

All these cases pose one interesting phenomenological question as to how
large $\theta _{\mu \nu }$ could be without contradicting existing
experimental data. To answer this question we have to adopt certain
calculational framework, incorporating together Standard Model fields and $%
\theta _{\mu \nu }$ background. The simplest way of doing this is to assume
that SM is realized on the brane and the external background of $B_{\mu \nu }
$ (or $\theta _{\mu \nu }$) field is included via Moyal product. 
In this
paper we take space and time independent background which could be a bad
approximation at large distances. Indeed, due to the interaction on the
brane, $B _{\mu \nu }(x)$ could become a massive field so that the
minimum of energy correspond to the vanishing vev. In this case the
constraints on $\theta _{\mu \nu }$ that we are aiming to
produce will be trivially satisfied since there is no $\theta$ to begin with 
and the theory is the usual one. Despite this possibility, we believe
that the question of experimental constraints on $B_{\mu \nu }$
 background/noncommutativity parameter $\theta_{\mu\nu}$ deserves special 
investigation.

Our strategy for the rest of this paper is quite straightforward. We take
the (*)-modified Standard Model and expand it once in the external $\theta $
parameter, 
\begin{equation}
f*g=fg+\frac{i}{2}\theta _{\mu \nu }\partial _{\mu }f\partial _{\nu }g,
\end{equation}
so that SM is extended by the series of dimension 6 operators, composed from
three or more fields: 
\begin{equation}
SM(*)=SM+\sum_{i}\theta _{\mu \nu }O_{\mu \nu }^{(i)}.  \label{sm*}
\end{equation}
Here the summation is performed over different types of operators, full list
of which is outside the scope of the present paper. First order in $\theta $
is sufficient for our purposes. Such a procedure of constructing an
effective action does not depend on taking the zero slope limit and holds
for a generic situation \cite{SW99}.

It is also to our advantage that in the effective Lagrangian approach we do
not have to worry about the renormalizability of this theory. Although the
ultimate resolution to the question of renormalizability of non-commutative
field theories is very interesting and very important \cite{renorm}, 
here we can simply
assume that all loop divergences are regularized at momenta comparable to $%
M_{s}$. Moreover, since the natural scale for $\theta$ is $ M_{s}^{-2}$, 
the iteration of $%
\theta _{\mu \nu }O_{\mu \nu }$-interactions does not create any problems as
long as external momenta are much smaller than $\sqrt{1/\theta }$.

\section{Experimental limits on $\theta_{\mu\nu}$}

{\em High-energy limits }

At first glance, it is advantageous to use high-energy processes to
obtain the most stringent limits on $\theta $. Indeed, the higher the
energy/momentum transfer is, the larger the effect of $O_{\mu \nu }$ will
be. Let us consider such a well-studied process as the Z-boson decay. We
shall profit from the fact that $O_{\mu \nu }$ violate Lorentz invariance
and use a decay channel which is strictly 
forbidden at the SM level and allowed
when $\theta _{\mu \nu }\neq 0$. A good candidate for this channel will be
the decay of Z into a pair of photons, forbidden by Lorentz invariance and
Bose statistics in SM. The experimental limit on the branching ratio for 
this decay is also very good, 
$Br(Z\rightarrow \gamma \gamma )\leq 5\cdot 10^{-5}$ \cite{PDG}.

To calculate this decay width we need to evaluate the 
$\tmn O_{\mu\nu}$-expansion of the $U(1)$ gauge sector, which
in the presence of external $\theta _{\mu \nu }$-background
could be taken in the following form: 
\begin{equation}
S_{nc}=-\frac{1}{4}\int d^{4}xF^{nc}_{\mu\nu}*F^{nc,\mu\nu}  \label{u1}
\end{equation}
Here $F_{nc}$ denotes ``non-commutative'' field strength given by 
\begin{equation}
F^{nc}_{\mu\nu}=\partial _{\mu }B_{\nu }-
\partial _{\nu }B_{\mu }-g^{\prime }\theta
^{\alpha \beta }\partial _{\alpha }B_{\mu }\partial _{\beta }B_{\nu }.
\label{fnc}
\end{equation}
Expanding Eq. (\ref{u1}) to first order in $\theta _{\mu \nu }$ we get 
\begin{equation}
S_{nc}=S+{1\over 2}g^{\prime }\int d^{4}x\theta ^{\alpha \beta }(\partial _{\mu
}B_{\nu }-\partial _{\nu }B_{\mu })\partial _{\alpha }B^{\mu }\partial
_{\beta }B^{\nu }.
\end{equation}
Going to the physical basis, we obtain the $Z\gamma \gamma $ interaction term: 
\begin{equation}
S_{int}={1\over 2}g^{^{\prime }}\sin \theta _{W}\cos ^{2}\theta _{W}\int 
d^{4}x\theta
_{\rho \sigma }[\partial _{\mu }Z_{\nu }\partial _{\rho }A_{\mu }\partial
_{\sigma }A_{\nu }+(\partial _{\mu }A_{\nu }-\partial _{\nu }A_{\mu
})\partial _{\rho }Z_{\mu }\partial _{\sigma }A_{\nu }].  \label{int}
\end{equation}
It should be noted here that the expansion of the $SU(2)$ sector 
has quadratic terms in $\theta _{\mu \nu }$, but not linear ones
and thus does not contribute into (\ref{int}).

Finally, we arrive at the following answer for the two-photon decay width of
the Z-boson, induced by the $\theta _{\mu \nu }$-background: 
\begin{equation}
\Gamma _{Z\rightarrow \gamma \gamma }={\frac{\alpha }{144}}\cos ^{4}\theta
_{W}M_{Z}^{5}\sum_{i}\theta _{0i}^{2}.
\end{equation}
The decay width is evaluated in the Lorentz system in which Z is produced at
rest.

Comparing this result with the experimental limits on the branching ratio,
we conclude that the sensitivity to $1/\sqrt{\theta }$ is not better than
250 GeV. This is a very modest limit, which could perhaps be improved had we
considered $\theta $-induced corrections to other high-energy processes such
as $e^{+}e^{-}$ cross sections, forward-backward asymmetry and so on. At any
rate, a significant improvement beyond the level of 250 GeV is not possible.
Thus, we conclude that the high-energy processes cannot produce sufficiently
strong bounds on $\theta _{\mu \nu }$, and turn to the low-energy limits
on this parameter.

\vspace{0.5cm}

{\em Low-energy limits }

The most notable feature of the effective Lagrangian (\ref{sm*}) is the
explicit violation of Lorentz invariance \cite{DS}. 
This violation is, of course,
controlled by the size of $\theta_{\mu\nu}$ and could be made arbitrarily
small ``by hand''.

The extension of Standard Model by Lorentz-noninvariant operators of
dimension $\leq $ 4, has been actively studied in the past \cite{Kost}. Some
of the limits obtained in \cite{Kost} are extremely strong. A priori, 
there are several
problems with such a generic description. If we believe that the Lorentz
non-invariant terms originate from short distances, it is not clear why
dimension 3 and 4 operators should be suppressed. Another problem is of
rather technical nature, as the number of free parameters in such extensions
of SM is generally very large. These problems do not exist in our
approach. The breaking of Lorentz invariance is given only by $\theta _{\mu
\nu }$, and its effect first shows up in dimension 6 operators.

The qualitative understanding of the role of $\theta _{\mu \nu }$ for
low-energy physics comes from the considerations of the non-relativistic
limit for the $\int d^4x~e\overline{\psi }*A_{\mu }\gamma ^{\mu }*\psi $ 
interaction
term taken in the external Coulomb field:
\begin{equation}
V=-{\frac{Z\alpha }{r}}-\frac{Z\alpha }{2r^{3}}({\bf \theta }_{B}\cdot {\bf L%
})-\frac{m_{e}Z\alpha }{r^{3}}({\bf \theta }_{E}\cdot {\bf r}).  \label{thL}
\end{equation}
Here we split $\theta _{\mu \nu }$ into two three-vectors in analogy with the
electromagnetic field, $\epsilon _{ijk}({\bf \theta }_{B})_{k}\equiv \theta
_{ij}$, $({\bf \theta }_{E})_{i}\equiv \theta _{0i}$. The diagonal matrix
element of the last operator in Eq. (\ref{thL}), taken over a wave
function of the discrete spectrum, is equal to zero, unless CP is broken. The
second term, however, may produce interesting phenomena, as it gives an
effective coupling of the angular momentum with an external vector ${\bf %
\theta }_{B}$, 
\begin{equation}
V=\kappa ({\bf \theta }_{B}\cdot {\bf J}).
\end{equation}
The effective coupling constant $\kappa $ has obvious $L,J$ dependence and, 
more importantly, is determined by the third power of the characteristic atomic
momentum. For an outer atomic electron this coupling is of the order of $%
Z^{2}\alpha (m_{e}\alpha )^{3}$. The coupling to $\theta _{B}$ creates a
Zeeman-like splitting of atomic orbitals with respect to an external vector
and thus can be tested in
high-precision atomic experiments.

At this point it becomes clear that similar effects (${\bf \theta }_{B}\cdot 
$ angular momentum) will appear in the hadronic physics, when we consider
the (*)-extended interaction of quarks and gluons. This interaction will lead to
the effective coupling of the nucleon spin with $\theta $,
\begin{equation}
\langle N|{\cal L}_{QCD}(*)|N\rangle =~
{\rm \theta\!-\!independent~terms}~~+~~\frac{d_{\theta }}{2}\theta _{\mu \nu }
\overline{N}\sigma _{\mu \nu }N,
\end{equation}
which in non-relativistic limit simply becomes  $d_{\theta }({\bf \theta }
_{B}\cdot \frac{{\bf S}}{S})$. The size of $d_{\theta }$ is given by a cube
of a characteristic hadronic scale. 

To estimate $d_{\theta }$ and to convince ourselves that such an effect
exists, we perform the following exercise, a simplified version of the nucleon
three-point function QCD sum rules \cite{Nsr}. 
We calculate the operator product
expansion (OPE) of two nucleon currents in the presence of the external $
\theta _{\mu \nu }$-background and observe the non-zero result. 
Then we take the
''phenomenological'' part of the QCD sum rule and saturate it with the
nucleon double-pole contribution. Matching the two sides at 1 GeV, we obtain an
estimate of $d_{\theta }$ for nucleons.

On the OPE\ side we can use an asymptotically free description, and thus
include the $\theta _{\mu \nu }$-piece as the correction to a free massless
quark propagator:
\begin{equation}
S(x,0)=\frac{ix_{\mu }\gamma _{\mu }}{2\pi ^{2}x^{4}}-\frac{ix_{\mu }\gamma
_{\beta }}{2\pi ^{2}x^{4}}t^{a}G_{\nu \beta }^{a}\theta _{\mu \nu }.
\end{equation}
This correction originates from the (*)-extended quark-gluon interaction, 
whereas
pure gluonic sector does not contain terms linear in theta. Using this
propagator it is straightforward to calculate the OPE of two nucleon
currents, expressing the result as the combination of a pure perturbative
piece  and non-perturbative condensates such as $\langle \overline{q}%
\sigma _{\mu \nu }q\rangle _{\theta _{\mu \nu }}$, $\langle \overline{q}%
\sigma _{\mu \nu }t^{a}G_{\mu \nu }^{a}q\rangle $, and so on. Deferring 
further details for more extended publication, we present here the final 
estimate
for $d_{\theta }:$%
\begin{equation}
d_{\theta }\sim 0.1{\rm GeV}^{3}
\end{equation}
Further progress in refining this estimate can be achieved by including the
anomalous dimensions, and fixing the size of ''susceptibility'' condensate $
\langle \overline{q}\sigma _{\mu \nu }q\rangle _{\theta _{\mu \nu }}$ by
comparing the stability of different sum rule channels.

As we remarked earlier, the  coupling of the nucleon spin to 
an external vector will lead to the
Zeeman-like splitting of the hyperfine structure in atoms. Since the
precision in measuring hyperfine splittings is at the level of mHz, this
should lead to a strong bound on $\theta _{B}.$ One of the most sensitive
systems in this respect is the hydrogen maser, where the effects of Lorentz
violating terms have been searched for in a recent experiment \cite{Harv}.
In this double-resonance experiment the absence of sidereal variations of
the maser frequency leads to 1 mHz limit on the product of $d_{\theta }$
and $\theta _{B\perp}$, component of $\theta _{B}$ perpendicular to earth's 
axis. This puts the limit on the magnetic component of $\tmn$ at the level 
$1/\sqrt\theta > 5 \cdot 10^{12}$ GeV. 

Another even more stringent limit could be extracted from the experiment 
which compares magnetic field measured by Cs and Hg atoms \cite{Amh}. 
The non-vanishing $\tmn$-background affects primarily nuclear spin. Thus the 
magnetic field, measured by mercury atom, will be corrected due to the 
interaction of the nuclear spin with $\theta_B$ to a larger extent than the 
magnetic field measured by Cs. The absence of sidereal variations 
in the difference of magnetic field measured by Cs and Hg is verified at
100 nHz level. This translates to the following, extremely tight bound
on $\theta$:
\be
\fr{1}{\sqrt\theta} > 5\cdot 10^{14}~ {\rm GeV}
\label{limit}
\ee
This is the main result of the present paper.

\section{Discussion}

The limit on $\theta_{\mu\nu}$ obtained in this paper is quite strong,
signaling the sensitivity of low-energy experiments to the scales
comparable to standard GUT/string scales. Of course, this sensitivity is the
consequence of the assumption about non-vanishing $\theta_{\mu\nu}$-dependent 
background, which
breaks explicitly Lorentz invariance.

Limit (\ref{limit}) is derived from the atomic 
experiments, performed in a laboratory. 
It would be  interesting to explore whether similar or stronger bounds
could be obtained from non-observation of the anisotropy of the Universe 
at large scales. Indeed, the coupling of baryon spins to 
$\theta _{\mu \nu }$ at
certain level should lead to the anisotropy of matter distribution and
polarization of interstellar medium which could, in principle, result in 
strong bounds on $\theta_{\mu \nu }$ \cite{CFJ}.

Returning to the brane-world scenario it is fair to question 
a ``natural'' value for $\theta_{\mu\nu}$ on the brane. For the
energies much higher than the inverse radius of extra dimensions, a natural
``string'' value of $\theta_{\mu\nu}$ is given by 
the square of the inverse string scale.
\footnote{
Throughout our discussion we {\em do not} assume zero slope limit, so that $%
M_{s}$ is the only natural dimensional high-energy parameter.} 
For $M_{s}\sim $ 1 TeV
this is in apparent contradiction with the limit (\ref{limit}).

Does our result (\ref{limit}) pose another ``naturalness'' problem for the
low-energy string scale models? The answer to this question depends on the
dynamical properties of $\theta _{\mu \nu }(x,y)$, where $y$ indicates the 
dependence on extra space-like coordinates. If the lowest mode of the 
Kaluza-Klein expansion of $\theta _{\mu \nu }(x,y)$
remains essentially massless, as it would happen 
in theories with extra dimensions of
infinite volume \cite{Rub}, then the vev of $\theta _{\mu \nu }(x,y)$ may
freeze at a large value, which would be in contradiction to the
observational evidence for absence of $\theta _{\mu\nu}$ at the level (\ref
{limit}). When the volume of extra dimensions is finite, it may turn out
that the ultimate value of $\theta _{\mu \nu }(x)$ is zero. Indeed, every
Kaluza-Klein copy of this tensor field may receive a mass term due to the
interaction with brane fields and/or details of mechanisms responsible for
compactification. The mass term induced by the interaction 
$\theta _{\mu \nu }O_{\mu \nu }$ on the brane could be estimated 
when the volume of extra
dimensions is large. If  we assume that the correlator of two $O_{\mu \nu }$
currents is saturated at the string scale $M_{s}$, 
\begin{equation}
\int \langle O(x),O(0)\rangle d^{4}x\sim ({\rm two~loop~factor})\times
M_{s}^{4},
\end{equation}
and the two-loop phase space factor originates from the fact that by construction 
every $O_{\mu \nu }$ has at least three fields. The kinetic term for
every Kaluza-Klein mode of $\theta _{\mu \nu }(x,y)$ receives a volume
enhancement factor, so that the effective mass for the lowest mode,
canonically normalized in four dimensions is 
\begin{equation}
m_{eff}^{2}\sim ({\rm two~loop~factor})\times {\frac{M_{s}^{2}}{%
M_{s}^{n}V_{n}}}\sim 10^{-4}\times {\frac{M_{s}^{4}}{M_{{\rm Pl}}^{2}}}\sim
(10^{-5}{\rm eV})^{2}\frac{M_{s}^{2}}{{1{\rm TeV^{2}}}}
\end{equation}
It is clear that for a TeV range $M_{s}^{2}$, $m_{eff}$ is very low, which
could create cosmological problems, similar to those arising in the
standard axion relaxation mechanism when the axion mass is small. In this
case the ``$\theta _{\mu \nu }$-problem'' is replaced by a new moduli problem
which indicates that the choice $M_{s}\sim 1$ TeV is unnatural. However,
the ultimate answer to the question about the mass of $\theta _{\mu \nu }(x)$
and its vev, surviving until present cosmological times, needs other physical
inputs such as information about the compactification mechanisms and 
supersymmetry breaking.

\bigskip {\bf Acknowledgments} M.P. thanks A. Lossev for very helpful
discussions and interest taken in this work. I.M. and R.R. thank I. Chepelev
for discussions. This work was 
supported in part by the Department of Energy under 
Grant No. DE-FG02-94ER40823 and NSF grant PHY-9722101.

\end{document}